# Emergent Tetragonality in a Fundamentally Orthorhombic Material

Anisha G. Singh[1,2,3*], Maja D. Bachmann[1,2,3], Joshua J. Sanchez[4], Akshat Pandey[5], Aharon Kapitulnik[1,2,3,5], Jong Woo Kim[6], Philip J. Ryan[6], Steven A. Kivelson[1,2,5,7], Ian R. Fisher[1,2,3]

[1] Geballe Laboratory for Advanced Materials, Stanford University; Stanford, CA, USA.
[2] Stanford Institute for Materials and Energy Sciences, SLAC; Menlo Park, CA, USA.
[3] Department of Applied Physics, Stanford University; Stanford, CA, USA.
[4] Department of Physics, Massachusetts Institute of Technology; Cambridge, MA, USA.
[5] Department of Physics, Stanford University; Stanford, CA, USA.
[6] Advanced Photon Source, Argonne National Lab; Lemont, IL, USA.
[7] Rudolf Peierls Centre for Theoretical Physics, University of Oxford; Oxford, UK.

**Abstract:** Symmetry plays a key role in determining the physical properties of materials. By Neumann's principle, the properties of a material remain invariant under the symmetry operations of the space group to which the material belongs. Continuous phase transitions are associated with a spontaneous reduction in symmetry. Much less common are examples where proximity to a continuous phase transition leads to an increase in symmetry. Here, we find signatures of an emergent tetragonal symmetry close to a charge density wave (CDW) bicritical point in a fundamentally orthorhombic material, $ErTe_3$, for which the two distinct CDW phase transitions are tuned via anisotropic strain. We first establish that tension along the a-axis favors an abrupt rotation of the CDW wavevector from the c-axis to the a-axis, and infer the presence of a bicritical point where the two continuous phase transitions meet. We then observe a divergence of the nematic elastoresistivity approaching this putative bicritical point, indicating an emergent tetragonality in the critical behavior.

**Introduction:**

The symmetry of a material prescribes much more than just the lattice vectors of a Bravais lattice. Tetragonal symmetry, for example, implies invariance of the crystal structure, and hence crystal properties, under 90 degree rotations about the principle axis, and certain other mirror and rotation symmetries depending on the space group. ErTe$_3$, like other rare-earth tritellurides (RTe$_3$), has a lower symmetry, belonging to the orthorhombic space group *Cmcm* ($D_{2h}^{17}$, no 63). The structure comprises bilayers of almost-square tellurium nets, which are separated along the *b*-axis by RTe block layers. At room temperature, the *a* and *c* lattice parameters are almost equal, $a = 0.999c$, and moreover, it is possible to tune the material from *a* > *c* to *a* < *c* using externally applied anisotropic strain. However, despite the near equivalence of the in-plane lattice parameters, the presence of a glide plane between the tellurium bilayers (Fig. 1A) makes the material fundamentally orthorhombic. The glide plane is a non-symmorphic symmetry element and cannot be removed by external strains. Thus, even when the material is strain-tuned to a point where the in-plane lattice parameters are exactly equal ($a = c$), the system does not possess a 4-fold rotational symmetry and hence can never be truly tetragonal. Nevertheless, as we find here, the presence of a strain-tuned CDW bicritical point in ErTe$_3$, which occurs at a critical strain where a ≠ c, yields signatures in the elastoresistivity of an emergent tetragonal symmetry associated with the critical fluctuations.

RTe$_3$ crystals host a unidirectional, incommensurate CDW. For un-strained ErTe$_3$, this onsets via a continuous phase transition at a critical temperature $T_{CDW}$ = 268K, with the wavevector oriented within the Te plane, along the *c*-axis (*1,2*). The instability is driven, at least in large part, by a strongly q-dependent electron-phonon coupling. Softening of the associated phonon mode as $T \to T_{CDW}$ has been observed via inelastic x-ray scattering (*3,4*). The structural motifs that make up the building blocks of RTe$_3$ result in a near-equivalence of the electronic structure and phonon spectrum in the two in-plane directions. This near-equivalence is reflected in a simultaneous, but incomplete, softening of the phonon mode at a wavevector equal in magnitude to the CDW wavevector but oriented in the transverse (*a*-axis) in-plane direction. Previous calculations indicate that unidirectional order is favored for sufficiently strong electron-phonon coupling, even on a truly square lattice (*5*). Consequently, the inequivalence of the *a* and *c* axes in ErTe$_3$ has generally been thought to act as a weak symmetry breaking field, favoring the *c*-axis CDW state over the competing *a*-axis state.

The presence of CDW fluctuations along both the *a* and *c*-axis directions (*3*), as well as the subsequent *a*-axis transition in ErTe$_3$ at a lower temperature (*6*), suggests it is possible to tune between these two CDW states using anisotropic strain. The CDW order parameters $\phi_a$ and $\phi_c$ couple to normal strains ($\epsilon_{xx}$, $\epsilon_{yy}$ and $\epsilon_{zz}$) as $\lambda_i^{jj}\epsilon_{jj}|\phi_i|^2$ where $\lambda_i^{jj}$ are coupling constants. There are two immediate consequences of this: First, the material develops a spontaneous anisotropic strain when cooled through the CDW transition (i.e., a nonzero value of $\phi_c$ or $\phi_a$ will result in a non-zero value of $\epsilon_{xx}$, $\epsilon_{yy}$, and $\epsilon_{zz}$). This is indeed observed: the lattice grows longer in the CDW wavevector direction and shorter in the transverse direction (*6*). Second, T$_{CDW}$ can be tuned by strain (*7*).

No symmetry operations of the point group relate the various coefficients $\lambda_i^{jj}$. However, the near equivalence of the electronic and phonon structures in the two directions suggests that certain coefficients might be closely related. For example, it is not unreasonable to anticipate that $\lambda_c^{xx} \sim \lambda_a^{yy} \sim -\lambda_a^{xx}$ etc. If this is indeed the case, then relatively small strains would not only be able to tune $T_{CDW}$, but also rotate the direction of the primary CDW wave-vector from along the

**Fig. 1**

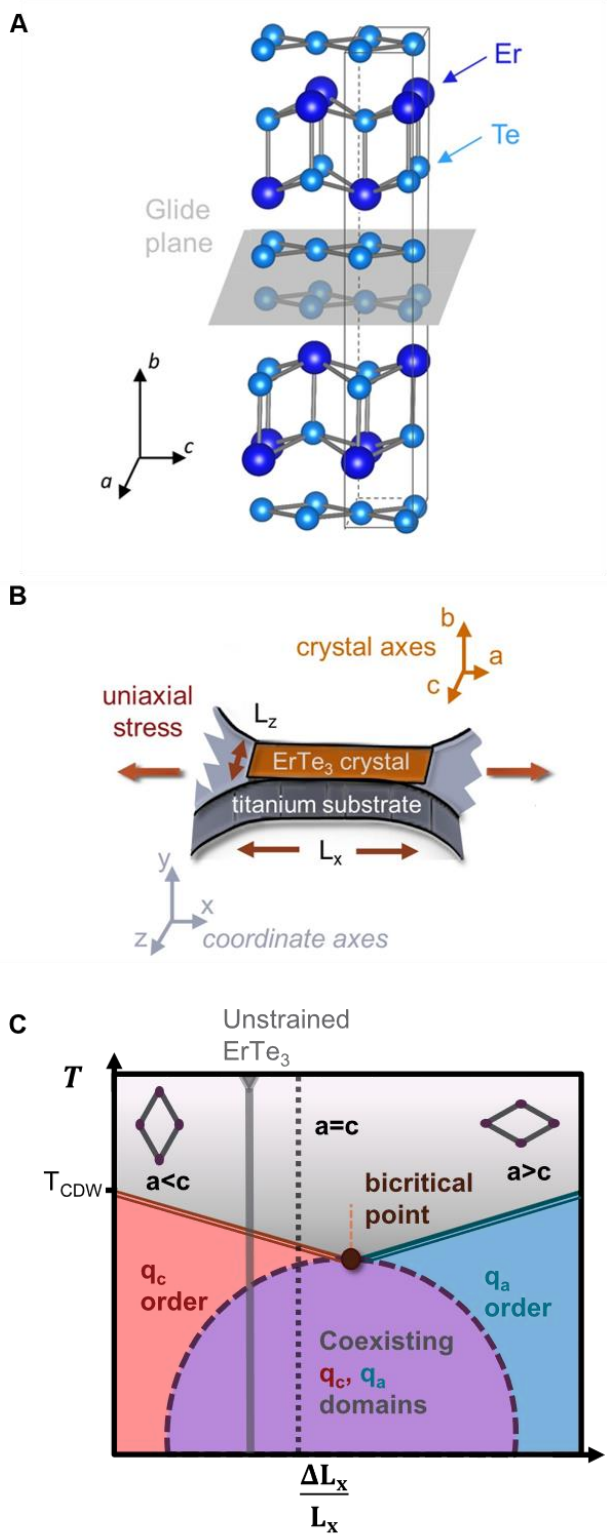

**Overview of uniaxial strain measurements of ErTe$_3$**: (**A**) Crystal structure of ErTe$_3$, which comprises Te bilayers separated by ErTe blocks. Note that in the *Cmcm* space group setting the in-plane lattice parameters are *a* and *c*, while the long *b* axis is perpendicular to the Te planes. A glide plane between the Te bilayers (illustrated) makes the material fundamentally orthorhombic. (**B**) Schematic diagram illustrating the neck of the titanium 'bowtie' platform used for applying strain to a thin single crystal of ErTe$_3$ which is bonded to its surface. Insets indicate coordinate axes for the experiment (xyz) and the crystal axes of the sample (abc). Labels indicate the effective length $L_x$ and width $L_z$ of the platform. (**C**) Schematic diagram illustrating the proposed phase diagram of ErTe$_3$ as a function of $\Delta L_x/L_x$. Double solid lines represent continuous phase transitions while dashed lines indicate first order phase transitions. Insets indicate deformation of the Te lattice as a result of the strain. The putative bicritical point is labeled.

c-axis to along the a-axis. Previous elastoresistance and elastocaloric effect measurements point towards this possibility, though direct evidence has thus far been lacking (*7*).

It is important to note that while both $\phi_a$ and $\phi_c$ are superficially similar CDW states, they are in fact fundamentally distinct from a symmetry perspective, precisely because the crystal structure is different in these two directions. At a minimum, this means it is not given that strain should necessarily be able to rotate the CDW direction and this must be verified experimentally. This is demonstrated here by performing high-resolution x-ray diffraction experiments on samples held under continuously variable strain conditions. As shown schematically in Fig. 1C, the strain-tuned phase boundaries of the two distinct CDW states meet, within experimental uncertainty, at a bicritical point below which there is a first order transition between the two ordered states. Because the experiment is conducted at fixed strain, not stress, the first order transition results in a region of two-phase coexistence - i.e., domains of the two distinct CDW states - for intermediate values of the macroscopic strain $\Delta L_x/L_x$ and below the bicritical temperature. Noting that there are no symmetry-imposed reasons dictating the existence of such a multicritical point, it is intriguing to find signatures in elastoresistivity measurements of an emergent tetragonal symmetry near the strain-tuned bicritical point. Specifically, we observe the divergence of the nematic elastoresistivity approaching the bicritical point, even though the material is structurally orthorhombic at this (and all) values of the applied strain. Such a divergence is ordinarily associated with the presence of nematic fluctuations in a tetragonal material. Hence this suggests the critical degrees of freedom in ErTe3 acquire an emergent additional symmetry near the bicritical point. We discuss the origin of this effect using a simple mean field perspective.

**High Resolution X-ray Diffraction:**

Utilizing high resolution x-ray diffraction (XRD) at a synchrotron light source, the evolution of CDW superlattice peaks with externally induced strain can be observed, providing a direct probe of the behavior of the CDW state. XRD measurements were completed at the Advanced Photon Source at Argonne National Lab at Sector 6-ID-B. Here, a sample environment has been developed to study samples with XRD with in-situ strain tuning at cryogenic temperatures (*8,9*). A CS-130 Razorbill strain cell was utilized to apply variable strain to the samples (*10*). Data was taken at 11.2 keV. For ErTe3, this corresponds to a penetration depth of roughly 50μm, which is greater than the thickness of samples studied.

RTe3 are very micaceous crystals. To maintain the crystallinity of the sample necessary for high resolution XRD while strain-tuning, rather than mounting the sample directly into the Razorbill cell, the sample was affixed to a titanium 'bowtie' platform designed for high strain transmission shown in Fig. 1B (*11*). The sample is less stiff than the Ti platform, and therefore deforms together with the platform when a stress is applied. In our measurement configuration, the *a*-axis was oriented in the stress direction (Fig 1B), resulting in an anisotropic strain state, with $\varepsilon_{xx}$ and $\varepsilon_{zz}$ related by the Poisson ratio of the Ti platform.

The strain-dependence of two CDW superlattice peaks are shown in Fig. 2A, confirming that anisotropic strain indeed reorients the primary CDW wavevector. In the absence of strain, the superlattice peak is observed along the *L* direction consistent with *c*-axis CDW order. As tension is applied along the *a*-axis, effectively reversing the anisotropy of the system, the *L* axis peak is suppressed while the *H* axis peak emerges, consistent with reorientation of the CDW wavevector by 90 degrees. Notably, reflective of the fundamental inequivalence of the in-plane

**Fig. 2**

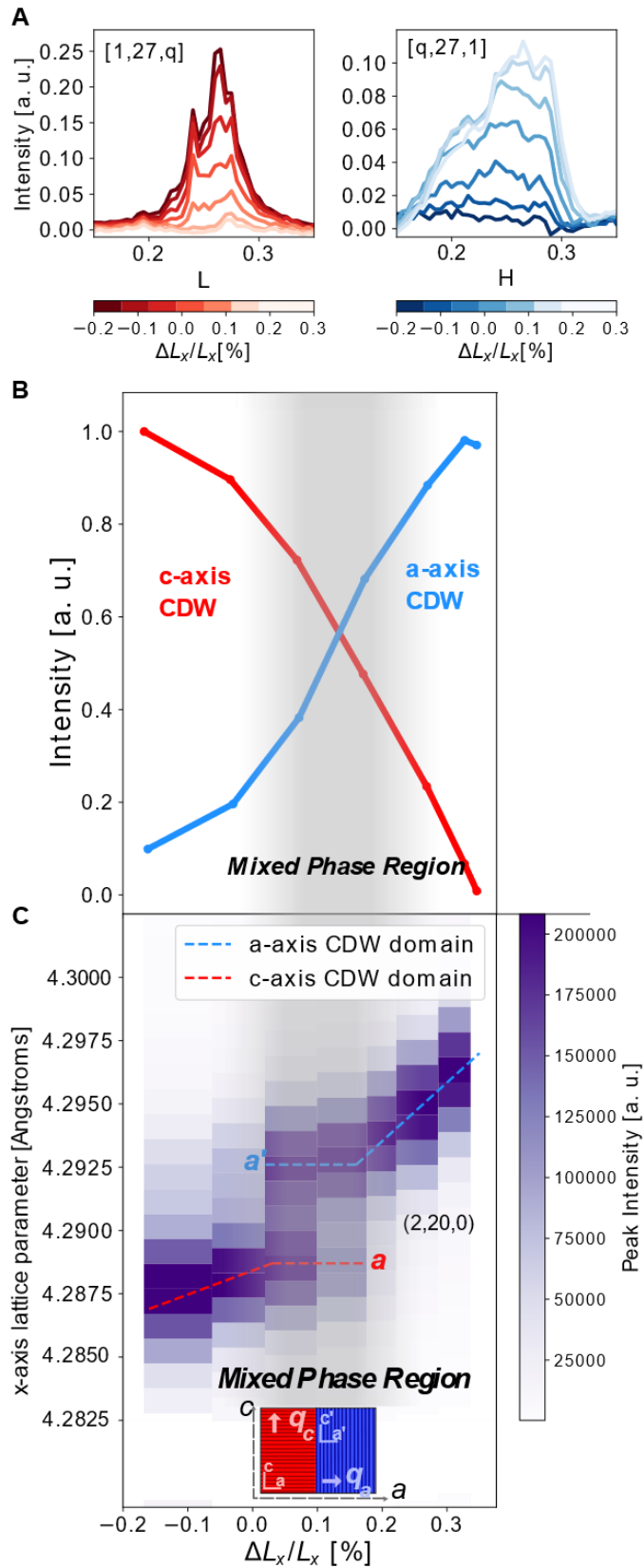

**XRD measurement of ErTe$_3$**: (**A**) Observed superlattice peak intensities at (1, 27, q$_{CDW}$) and (q$_{CDW}$, 27, 1) at 200K for different values of $\Delta L_x/L_x$. Under tension, the *c*-axis CDW peak is suppressed while the *a*-axis CDW peak increases in intensity. Substantial broadening is observed in these structural peaks relative to free-standing crystals due to sample manipulation in the strain cell. (**B**) Integrated intensity of CDW superlattice peaks shown in (A) after background subtraction and normalization to adjacent Bragg peak intensities. Shaded region indicates mixed $q_c$, $q_a$ phase. The mixed phase is defined here where neither CDW state exceeds 90% of the total CDW intensity. (**C**) 2D plot of the intensity of the (2,20,0) Bragg peak as a function of strain at 200K. For intermediate strains, two peaks appear in the data due to the presence of domains, corresponding to regions of the material where the CDW wavevector is oriented along the *a* and *c* axes respectively, as illustrated in the inset schematic. As consequence, the material effectively has two different lattice parameters along the x-axis here labeled *a* and *a'*. In this strain range, the in-plane lattice parameters are constant since the system accommodates applied strain by changing the relative population of the two CDW domains.

axes, additional measurements reveal that the magnitude of the wavevector of the rotated CDW is close to but distinct from the unrotated value (see Fig. S3). A similar rotation of the CDW wavevector with anisotropic strain was recently, independently observed for TbTe$_3$ (*12*). Our measurement also reveals a range of intermediate strains where the *c* and *a*-axis CDW are simultaneously observed.

In Fig. 2B, the integrated intensity of both superlattice peaks is plotted as a function of $\Delta L_x/L_x$. The total CDW intensity is constant with strain, but shifts between the *c* and *a* axis CDW states. In Fig. 2C, the behavior of a structural Bragg peak with an in-plane component, is plotted for the same strain range, reported here as the lattice parameter along the *x*-axis. Where the mixed CDW phase is observed, the Bragg peak splits, indicative of the emergence of domains. In this strain range, by rotating the CDW wavevector, the system adopts structural domains through which it can internally relax the applied stress. If, however, the applied strain exceeds the internal strain created by CDW rotation, a mono-domain state is again achieved and the lattice parameter varies monotonically with strain. Notably, the observation of such structural domains provides clear evidence that the simultaneous observation of $q_a$ and $q_c$ order is not the consequence of the emergence of a uniform homogeneous state comprising both wavevectors. Moreover, no peak was observed at the $q_a + q_c$ wavevector.

**Elastoresistivity:**

Transport measurements can provide a useful window on electronic behavior close to phase transitions, with important information encoded in both its strain and temperature dependence. Here, we focus on the in-plane resistivity anisotropy ($\rho_a - \rho_c$), which is sensitive to the orientation of the CDW wavevector.

To perform accurate measurements of the in-plane resistive and elastoresistive anisotropies, a unique transport device was designed. The device incorporates a transverse measurement geometry which allows for direct determination of the resistive anisotropy in a single measurement (i.e. the measured voltage from the transverse contacts is directly proportional to $\rho_a - \rho_c$; see refs *13,14*). Additional details of the device are presented in Fig. 3A and in the SM. The sample geometry was defined using a focused ion beam (FIB) and utilizes the same titanium platform that was used for the XRD measurements as shown in Fig. 3B.

Measured values of the in-plane resistive anisotropy are plotted in Fig. 3C. For temperatures above the onset of CDW order, $T > T_{CDW}$, $\rho_a - \rho_c$ is close to, but not exactly zero, consistent with the near equivalence of the two inequivalent crystallographic directions. Below the CDW transition, opening of the CDW gap results in an increase in the resistivity, though how this affects $\rho_a$ and $\rho_c$ depends on the orientation of the CDW wavevector. When the wavevector orients along the c-axis, there is a larger change in $\rho_a$ compared to $\rho_c$, resulting in a positive value of $\rho_a - \rho_c$ (*13*). This behavior has been previously understood using a simple Boltzmann transport approach including anisotropy in the Fermi velocity (*15*). Conversely, when the CDW orients along the a-axis, the change in $\rho_c$ is now larger than that of $\rho_a$ resulting in a negative value of $\rho_a - \rho_c$ (*7*).

The resistivity data also reveals a change in $T_{CDW}$ with strain. Taking the peak in the temperature derivative of the longitudinal resistivity ($\rho_a + \rho_c$)/2 as a measure of the critical temperature (see Fig. S4), $T_{CDW}$ can be determined as a function of strain, and is shown in Fig. 4A. This data reveals a minimum value of $T_{CDW}$ close to a critical value of $\Delta L_x/L_x \sim 0.13\%$. Since the data is plotted versus the macroscopic or applied strain rather than intrinsic sample strain as

**Fig. 3**

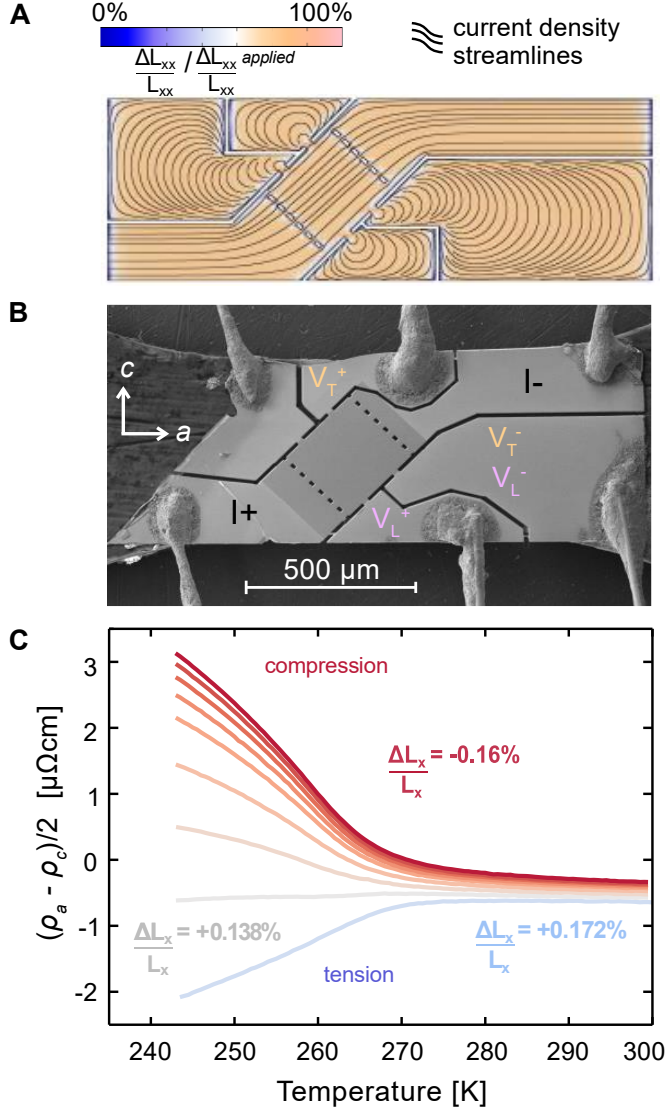

**Transport measurements in ErTe$_3$ under uniaxial strain**: (**A**) Finite element simulations, utilizing COMSOL Multiphysics, of the strain transmission and current flow for a typical transport device performed to establish optimal sample geometry to achieve large, homogeneous strain transmission as well as equipotentials that are closely parallel necessary for resistivity measurement. See SM for additional details (**B**) SEM image of a typical device measured. Signal measured between the VT +/− contacts, the transverse voltage, correspond to the in-plane resistive anisotropy ($\rho_a$–$\rho_c$)/2 (**C**) In-plane resistive anisotropy as a function of temperature for different (representative) values of compressive and tensile strain applied parallel to the a-axis of the sample.

experienced by each unit cell in the material, this minimum observed in $T_{CDW}$ is potentially broadened since the intrinsic sample strain changes minimally in this region of the plot due to domain formation. The overall shape of the resistivity curves does not change on tuning through this value of the strain, indicating that the phase transitions remain continuous. Notably, this is the same strain for which $\rho_a - \rho_c$ does not increase or decrease below $T_{CDW}$ (Fig. 3C), implying that $T_{CDW}$ at this strain value marks the putative bicritical point at which the onsets of $\phi_a$ and $\phi_c$ CDW order coincide. As discussed earlier, there is no symmetry relation between the *a* and *c*-axis CDW states. Consequently, the strain derivative of the critical temperature of the two states is not compelled to be the same, and is indeed found to be the case, i.e. $T_{CDW}$ is not an even function about the critical strain in Fig. 4A.

Inspection of the data shown in Fig. 3C also reveals that the largest changes in resistivity with respect to strain occur when the resistive anisotropy is smallest. To further investigate this, elastoresistance measurements were performed as a function of strain and temperature close to $T_{CDW}$. Utilizing an AC strain technique, in which a small AC strain is superimposed on an offset

DC strain and the resulting change in resistivity is measured using a demodulation technique (*16*), we can directly and sensitively measure the strain derivative of the resistive anisotropy. This value, normalized by the longitudinal resistivity, is plotted as a function of the relative temperature ($T - T_{CDW}$) in Fig. 4B, for strains below, at and above the critical strain. To determine the relative temperature, $T_{CDW}$ as a function of $\Delta L_x/L_x$ was found by fitting the data shown in Fig. 4A to a smooth polynomial. The elastoresistivity data shown in Fig. 4B clearly peaks upon approaching $T_{CDW}$ at the critical strain value (red data) but has a smaller magnitude and rolls over for all other values of applied strain (grey data).

The same elastoresistivity coefficient is plotted as a function of strain in Fig. 4C for different values of the relative temperature. This reveals a clear maximum centered around the critical strain 0.13%. This is a remarkably robust observation and is not sensitive to the exact functional form used for the strain dependence of $T_{CDW}$ (see Fig. S5).

**Discussion:**

The differential "nematic" elastoresistance

$$\eta = \frac{1}{(\rho_a + \rho_c)} \frac{\partial(\rho_a - \rho_c)}{\partial(\epsilon_{xx} - \epsilon_{zz})} \quad (1)$$

has no special meaning in an orthorhombic material. The three normal strains, $\epsilon_{xx}$, $\epsilon_{yy}$ and $\epsilon_{zz}$ all belong to $A_{1g}$ representations in an orthorhombic point group, such that each strain can separately affect each of the terms $\rho_a$, $\rho_b$ and $\rho_c$ in the resistivity tensor. Consequently, while this quantity is allowed to vary with strain and temperature, there are no symmetry constraints that would dictate singular behavior for this combination of coefficients.

In contrast, for a tetragonal material with a two component order parameter ($\phi_x$, $\phi_y$), this coefficient (rotating the coordinate axes so the $x$ and $y$ directions are related by a 4-fold symmetry) has a specific physical meaning. Now, a nematic order parameter can be defined: $N = <\phi_x^2> - <\phi_y^2>$, a quantity which, above the critical temperature, measures differences in the fluctuations of the two components of the order parameter. Since $N$ couples linearly to antisymmetric strain $\epsilon_{B1g} = (\epsilon_{xx} - \epsilon_{yy})/2$, a nematic susceptibility for the same symmetry channel, $\chi_{B1g} = \frac{\partial N}{\partial \epsilon_{B1g}}$, can be defined. In this instance, as has been extensively discussed in the context of Fe-based superconductors (*17-19*), the resistive anisotropy is linearly proportional to $N$ for small values of N, such that the nematic elastoresistivity coefficient η is proportional to χB1g (*18*). Therefore, a divergence in η in a tetragonal material is evidence of a diverging nematic susceptibility. Such behavior has been observed for a variety of materials for which multicomponent order parameters are relevant, including Fe-based superconductors and BaNi2As2 (*17,20*).

This behavior is not anticipated for ErTe3 since ϕa and ϕc do not belong to a multicomponent order parameter and a 'permanent' inequivalence of x and z directions means that N has a finite value for all temperatures and strains (barring accidental degeneracies). Such a system would not show any singular behavior in the elastoresistivity beyond standard Fisher-Langer-like anomalies at the strain-tuned phase transition (*21*). The sharp increase of η observed for ErTe3 approaching the putative bicritical point is therefore anomalous and is highly suggestive of an emergent tetragonal symmetry proximate to this point in the temperature-strain plane. We initially adopt a simple mean field model to provide insight into this remarkable observation. To quartic order in the two order parameters, and neglecting gradient terms and the elastic energy, the contribution to the free energy density from the CDW is given by:

**Fig. 4**

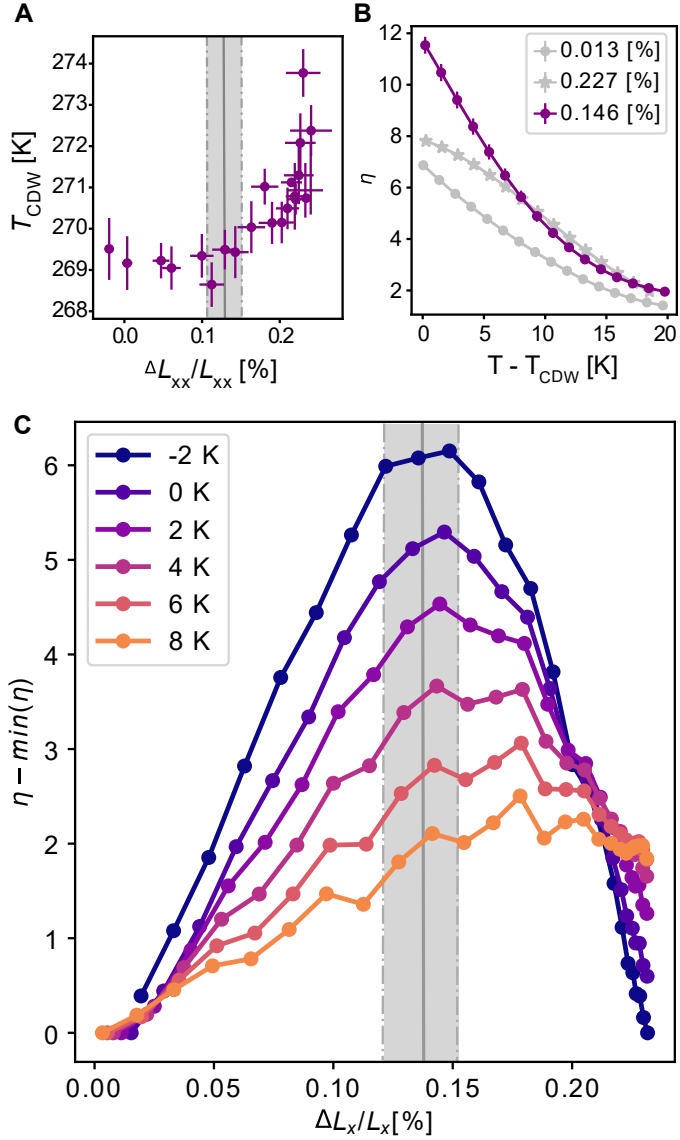

**Elastoresistivity measurements of ErTe$_3$:** (**A**) The strain dependence of T$_{CDW}$, as determined from the peak in the temperature derivative of the longitudinal resistivity $(\rho_a + \rho_c)/2$ (see Fig. S4) and expressed here as a function of $\Delta L_x/L_x$. Vertical band marks the critical strain at which the strain-derivative of T$_{CDW}$ changes sign and its uncertainty. (**B**) In-plane elastoresistivity anisotropy versus temperature for fixed offset strains (labelled in legend). For strain values close to the critical strain (red curve) this quantity diverges as T approaches T$_{CDW}$. While the strains in this plot are reported as a constant, this data was taken for a fixed voltage applied to the piezoelectric stack. The strain label is therefore an average over the temperature sweep with an uncertainty of +/-0.02%. For this reason, the strain at which a maximum response is observed in these temperature sweeps differs slightly from the maximum strain response reported in panel (C), where instead strain sweeps at constant temperature are reported. (**C**) In-plane elastoresistive anisotropy, η as defined in Section 4, versus strain for fixed relative temperature. The minimum value of η for a given relative temperature is subtracted from all data at that temperature to facilitate comparison across temperatures on a single plot. Vertical band marks the critical strain, and its uncertainty.

$$\Delta F = r_a|\phi_a|^2 + u_a|\phi_a|^4 + r_c|\phi_c|^2 + u_c|\phi_c|^4 + g|\phi_a|^2|\phi_c|^2 \quad (2)$$
$$+ \lambda_a^{xx} \epsilon_{xx}|\phi_a|^2 + \lambda_a^{zz} \epsilon_{zz}|\phi_a|^2 + \lambda_c^{xx} \epsilon_{xx}|\phi_c|^2 + \lambda_c^{zz} \epsilon_{zz}|\phi_c|^2$$

where $|\phi_a|$ and $|\phi_c|$ describe the amplitude of the two incommensurate CDW states. Here, to account for the fact that the competition between the phases is sufficiently strong to yield a bicritical point, we consider $g > 2\sqrt{u_a u_c}$. The strain coupling terms have been previously defined, and we neglect equivalent terms in $\epsilon_{yy}$ for simplicity. In the simplest possible phase diagram as a function of $\epsilon_{xx}$, $\epsilon_{zz}$ and temperature, the planes defining T$_{CDW}$ for each phase meet along a line of bicritical points (*22*). The experiment has a fixed relation between $\epsilon_{xx}$ and $\epsilon_{zz}$ defined by the Poisson ratio of the Ti platform, and thus we observe two lines which meet at a single bicritical point (Fig. 4A).

At the bicritical point, the coefficients of quadratic terms in the free energy, $r_a$ and $r_c$, have both been tuned to zero. Neglecting for now the strain terms, a simple redefinition of the units of either of the CDW order parameters can lead to an equivalence of the quartic terms, such that exchange of $\phi_a$ and $\phi_c$ leaves the free energy unchanged. (See SI for full derivation) In other words, to quartic order in $\phi_a$ and $\phi_c$, and neglecting other terms in the free energy, there is an emergent tetragonal symmetry associated with the critical degrees of freedom, which is inherited by all thermodynamic and transport properties sensitive to the critical fluctuations.

Even within this mean field picture, however, this is clearly only an approximate symmetry. Considering if the strain terms had a strict tetragonal symmetry (i.e. $\lambda^{xx}_c = \lambda^{zz}_a = -\lambda^{xx}_a$ etc.) then rotation of the x and z axes would leave the free energy invariant. However, allowing inequivalence of these terms, or including gradient terms or other higher-order terms, all lead to an inequivalence of the *a* and *c* crystal directions. Thus, the mean field anticipation is for an emergent *approximate* higher symmetry close to the strain-tuned bicritical point. To the level of precision of the present experiments, this mean field perspective is sufficient to account for the divergence of the nematic elastoresistivity coefficient $\eta$ upon approaching the putative bicritical point. Remarkably, a thorough treatment of the fluctuations, which we develop in a separate article, reveals that under certain circumstances, this emergent symmetry is in fact asymptotically exact upon approach to criticality (*22*), though it is unlikely that experiments with the present control of strain homogeneity could distinguish the associated proposed scaling behavior from the mean field expectations.

**Concluding Remarks:**

In closing, we comment on some wider implications of our observations. First, having established that modest strains can rotate the direction of the CDW in ErTe$_3$, we note that domain formation is to be anticipated whenever a crystal of RTe$_3$ is held under conditions that fix the lateral dimensions (if the spontaneous strain that develops at T$_{CDW}$ exceeds that which is caused by fixing the sample dimensions). In particular, one should anticipate domain formation whenever thin samples are bonded to a platform. Since this is a standard geometry for many experiments, consideration of domain formation should be included in interpretation of a variety of experimental results. Indeed, earlier STM results showing domain formation on the surface of TbTe$_3$ crystals might be related to this phenomenology (*23*).

Second, while the specific symmetry element that renders ErTe$_3$ orthorhombic for all values of externally induced strains is a glide plane, other structural motifs, for example the 1d chains in YBa$_2$Cu$_3$O$_{7-\delta}$, can play a similar role. The present observations provide proof of principle that similar phenomenology of emergent higher symmetries could, at least in principle, be found in such systems.

Finally, we note that these observations fall within a wider context of materials for which non-symmetry enforced degeneracies appear to play a crucial role in establishing emergent properties. Other examples include the possible multicomponent superconducting state in UTe$_2$, another fundamentally orthorhombic material for which degeneracy of two singlet representations has been proposed (*24*); possible multicomponent superconducting states in Fe-based superconductors (arising from 'accidental' degeneracy of singlet s and d-wave states in these tetragonal materials (*25, 26*); and proposals for other exotic non-symmetry-enforced multicomponent superconducting states in Sr$_2$RuO$_4$ (*27-29*).

**Supplementary Materials:**

Crystal Growth and Sample Preparation

Single crystals of $ErTe_3$ were synthesized via a self-flux method as described in (*1,2*). Typical crystals had a surface area between 1 -2 $mm^2$ and were 0.5 mm thick. Prior to mounting the sample for strain measurement, the crystal's in-plane axes must be distinguished to select the strain axis. This is accomplished via x-ray diffraction by utilizing the forbidden extinctions of the *Cmcm* space group. (0, *K*, *L*) peaks are permitted for even values of *K*, while (*H*, *K*, 0) peaks are only permitted if both *H* and *K* are even. Hence by comparing amplitudes of the (0, 6, 1), an allowed reflection, with (1, 6, 0), its forbidden counterpart, the a and c axes can be distinguished. No intensity was observed at the forbidden reflection, within the resolution of our measurement, indicating samples measured do not exhibit any stacking faults in which the a and c crystal axes are reversed.

Sample Mounting for Strain Measurement

The sample itself must be sufficiently thin to mitigate strain relaxation over its height. Finite element simulations indicated that the strain over the height of the sample is approximately homogeneous for crystals less than 50um thick, (Figure S1). $RTe_3$ crystals exfoliate easily due to weak van der Waals bonding between the tellurium sheets, making it straightforward to cleave samples to thin dimensions. Similarly, a thin and stiff glue layer is necessary to mitigate strain relaxation. For sample mounting, Angstrom Bond epoxy was used which yielded glue layers between 5-10um thick. Samples used for XRD measurement were between 10-20um thick, while samples used for transport measurements were between 1-10um thick. In this range of thicknesses, all samples behaved similarly with strain, however thinner samples were found to have worse crystallinity and hence were avoided for XRD measurement.

Device Details for Transport Measurement

In order to perform transport measurements, the titanium platform which was to be used for the experiments was initially heated to high temperatures for several hours to create an insulating oxide layer prior to sample bonding. To achieve high and homogeneous strain transmission in the transport device, the sample length along the strain direction should be maximized, resulting in the choice of a diamond shape for the active area, rather than more traditional bar shaped sample geometry. For this wider sample, multiple current injection points must then be defined in order to facilitate homogeneous current flow in the sample. Design was optimized using finite element simulations.

Measurement of Strains in the Experiment

In this experimental set-up, a uniaxial stress is applied to the sample platform producing strains $\Delta L_x/L_x$, $\Delta L_y/L_y$, and $\Delta L_z/L_z$ in the platform. $L_x$ is the effective or active length of the platform which for this "bowtie" design has been determined by simulation to be 3.47mm. The capacitive sensor in the strain cell allows for measurement of the displacement between the cell's jaws. For low strains we approximate that this displacement is entirely transmitted to the active (narrow) region of the 'bowtie' platform, giving a measurement of $\Delta L_x$. (In practice though, this transmission is likely between 80-90%). However, once the elastic limit of titanium is exceeded, at approximately 0.2%, the capacitive reading is no longer an accurate determination of $\Delta L_x$ since the shape of the platform has plastically deformed creating a substantial offset in the actual strain. Instead for large displacements, $\Delta L_x$ can be determined by mounting a strain gauge directly onto the Ti platform and a corrective function can be developed to convert the capacitance reading to the actual strain, as shown in Figure S2. This correction is the primary origin of the x-axis error bars shown in Figure. 4A.

While we report our data in terms of measured platform strain, this is a reasonable approximation for the macroscopic strain experienced by the ErTe$_3$ sample. Simulations of samples studied reveal that close to 90% of the platform strain $\Delta L_x/L_x$ is transmitted to the sample. This transmission can also be verified by direct measurement of the sample lattice parameters with XRD. The width of our samples was chosen such that the transverse strain $\Delta L_z/L_z$ is set by the Poisson ratio of the platform rather than the sample. The sample will experience additional strains however due to differential thermal contraction between it and the platform. Since RTe$_3$ contracts 5 times faster than titanium, the sample will experience additional tensile strains as it cools. This effect is not reflected in the reported value of strain.

The sample's macroscopic strain is different from its microscopic strain in the CDW state, as discussed in the main text. Above the CDW state, $\epsilon_{xx}$ and $\Delta L_x/L_x$ are equivalent. This changes with the presence of domains. While the dimensions of the sample change when the platform is strained, within a single domain $\epsilon_{xx}$ is zero and will not become nonzero until the sample is strained into one of the monodomain states. This can be seen clearly in Fig. 2B where there is no change in the lattice parameter in the mixed domain phase.

Determination of T$_{CDW}$

As discussed in the text, T$_{CDW}$ is determined by taking a temperature derivative of the longitudinal resistivity and then locating the peak in that derivative as demonstrated in Fig. S4. (This data is plotted inverted such that the feature corresponds to a peak rather than a dip). Data sets of both the raw data and derivative are offset for clarity. Additionally, only a representative subset of the extracted data shown in Figure 4A is illustrated here.

Elastoresistivity as a Function of Strain

The data presented in Figure 4C is influenced by the definition of the relative temperature. In Fig. S5, the elastoresistivity data are replotted three different ways. First, as done in Fig. 4C, a smooth function is used to model $T_{CDW}$ as a function of $(\Delta L_x/L_x)$ producing the most accurate result and appropriate even far from the bicritical point. Second, two lines are used to model $T_{CDW}$ as a function of $(\Delta L_x/L_x)$ producing a method that is less accurate over the entire strain range, but presumably is still a good description close to the bicritical point. Finally, $T_{CDW}$ is modeled as a constant, clearly an unphysical choice, but is included to demonstrate how robust the result is. Comparing Figures S4A and S4B, although the exact shape of the peak in the elastoresistance changes depending on the definition of the relative temperature, the peak feature itself is robust. If $T_{CDW}$ is assumed to be constant (Figure S4C), the peak feature is less pronounced, because the elastoresistivity response grows rapidly at high strains due to the strong enhancement of $T_{CDW}$ which is here neglected. Nevertheless, a peak in the elastoresistivity is still observed close to $T_{CDW}$ even in this case.

Tetragonal Approximation of Mean Field Model at Bicritical Point:
Starting from equation 2 in the main text:
$$\Delta F = r_a|\phi_a|^2 + u_a|\phi_a|^4 + r_c|\phi_c|^2 + u_c|\phi_c|^4 + g|\phi_a|^2|\phi_c|^2$$
$$+ \lambda_a^{xx}\epsilon_{xx}|\phi_a|^2 + \lambda_a^{zz}\epsilon_{zz}|\phi_a|^2 + \lambda_c^{xx}\epsilon_{xx}|\phi_c|^2 + \lambda_c^{zz}\epsilon_{zz}|\phi_c|^2 + ...$$

At the bicritical point, the coefficients of the quadratic order parameter terms will go to zero. The expression becomes:

$$\Delta F = u_a|\phi_a|^4 + u_c|\phi_c|^4 + g|\phi_a|^2|\phi_c|^2 + \ldots$$

At this point, we can define a renormalization $\phi_c = \left(\frac{u_a}{u_c}\right)^{\frac{1}{4}} \phi'_c$ such that the free energy expression can be rewritten as

$$\Delta F = u_a|\phi_a|^4 + u_a|\phi'_c|^4 + g'|\phi_a|^2|\phi'_c|^2 + \ldots$$

which is now invariant between switching of $\phi_a$ and $\phi'_c$.

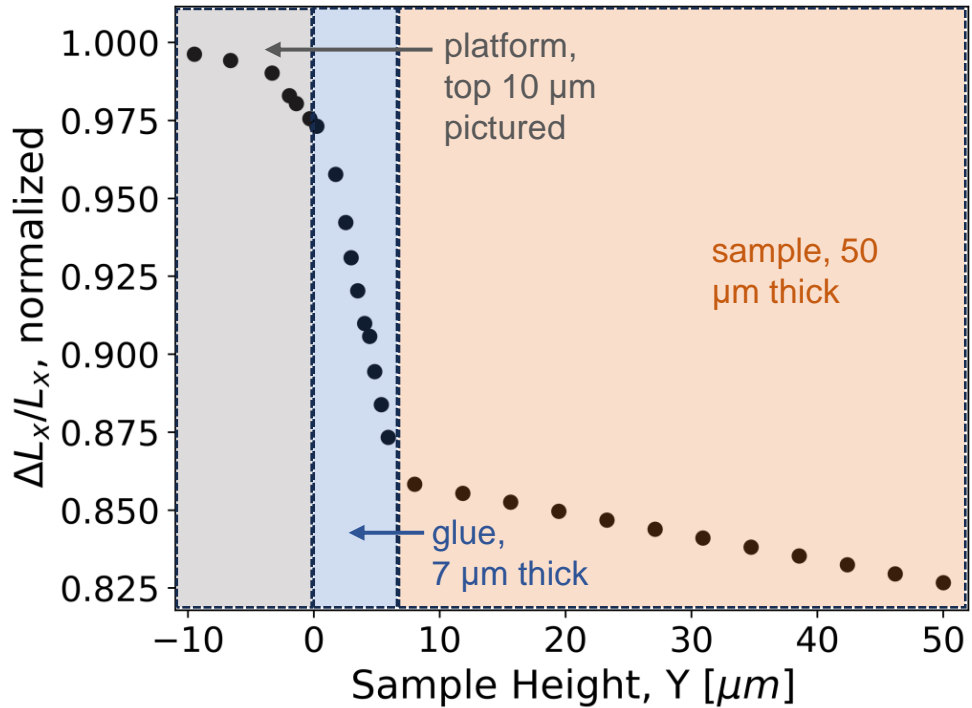

**Fig. S1.**
**Finite element analysis of sample strain across device height:** In this simulation, we model the behavior of the device neck with glue and an $ErTe_3$ sample affixed to it with thicknesses defined in the figure. The coordinate axes are defined such that Y= 0μm corresponds to the top surface of the titanium platform. In this simulation, one of the [y, z] faces of the titanium platform is held fixed while a prescribed displacement is then applied along the x- direction. The strains plotted are normalized with respect to this prescribed strain, in order to easily interpret how much the applied strain decays over the height of the device. There is some strain relaxation at the top surface of the platform due to the presence of the sample. The strain transmission falls off mostly through the glue layer, emphasizing the need for thin glue layers. In contrast, over the height of the $ErTe_3$ sample, the applied strain is relatively constant. For the 50 μm thick sample pictured, the variation in strain with sample height is several orders of magnitudes smaller compared to the strain variation applied in the experiment.

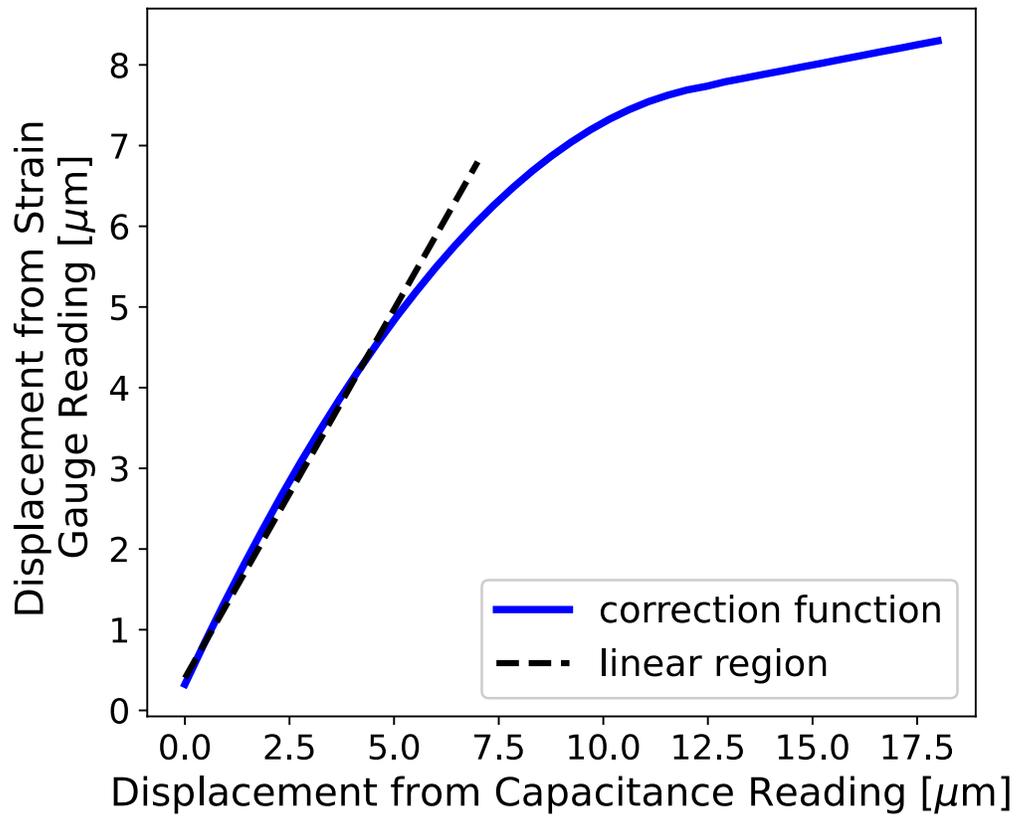

**Fig. S2.**
**Nonlinear response of platform at high offset strains:** Displacement measured from strain gauge on platform as a function of displacement measured from capacitive sensor shown in blue. At low strains, this function is linear (dashed black line). The strain gauge reading starts to deviate substantially from the capacitive reading at 7um.

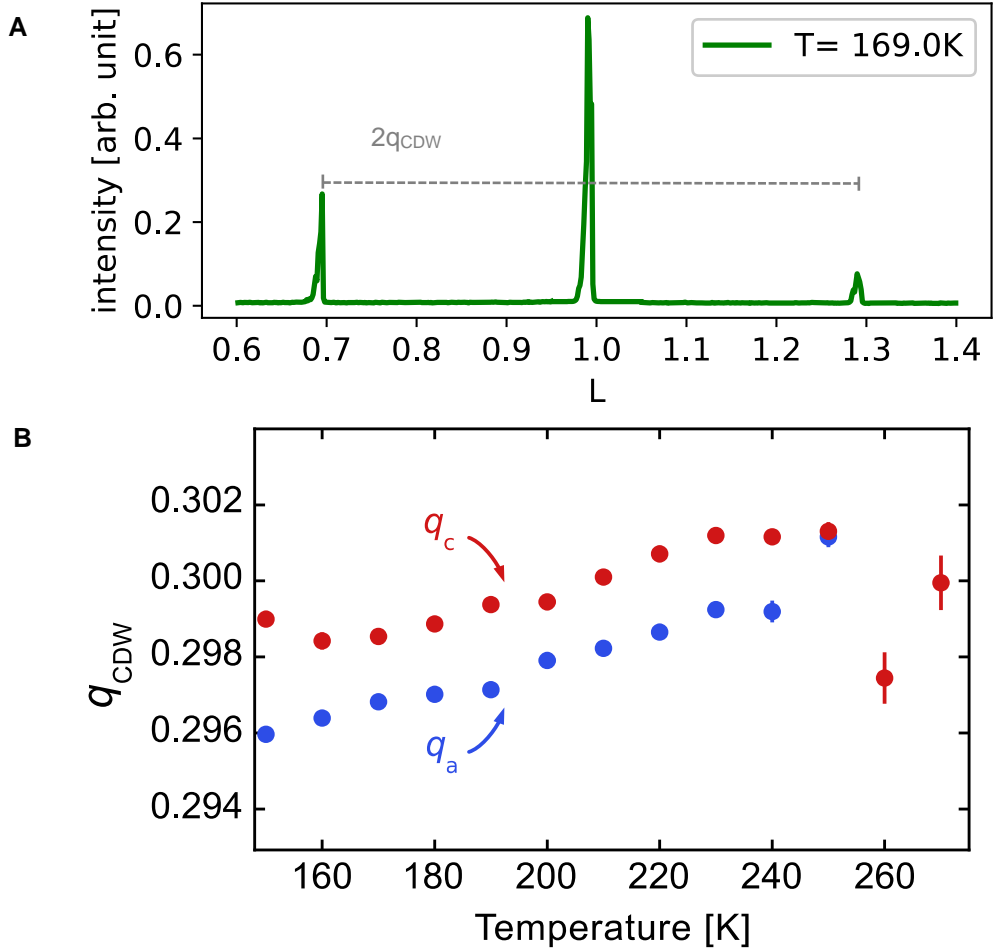

**Fig. S3.**

**q$_{CDW}$ wavevectors as a function of temperature:** A) Scan along L direction on (h, k) = 1, 27 plane. By taking the difference between the positions between a pair of superlattice peaks the value of the CDW wavevector can be determined as shown. B) Values of the CDW wavevector observed along the H and L directions specifically, the quantity q$_a$, plotted in blue, is the observed value of q$_{CDW}$ for the (q$_{CDW}$, 27, 1) peak in units of a* and the q$_c$ value, plotted in red, corresponds to the value of q$_{CDW}$ observed for the (1, 27 q$_{CDW}$) peak in units of c*. As an incommensurate order, q$_{CDW}$ can vary with temperature. Both the values of q$_a$ and q$_c$ are similar to the value of q$_{CDW}$, 0.298 c*, previously observed in ErTe$_3$ (*6*), but notably, the values of q$_a$ and q$_c$ are systematically different from each other. Data presented was collected on a sample mounted on a sapphire rather than titanium platform. In the dataset presented, no bias strain has been applied, but like the sample mounted on titanium, the sample still enters a mixed domain state as a consequence of differential thermal contraction. Since the sample is not manipulated or dynamically strained after bonding to the sapphire, in contrast to the data shown in Figure 2A, the CDW diffraction peaks here are much sharper such that the clear difference in the wave vector can be observed.

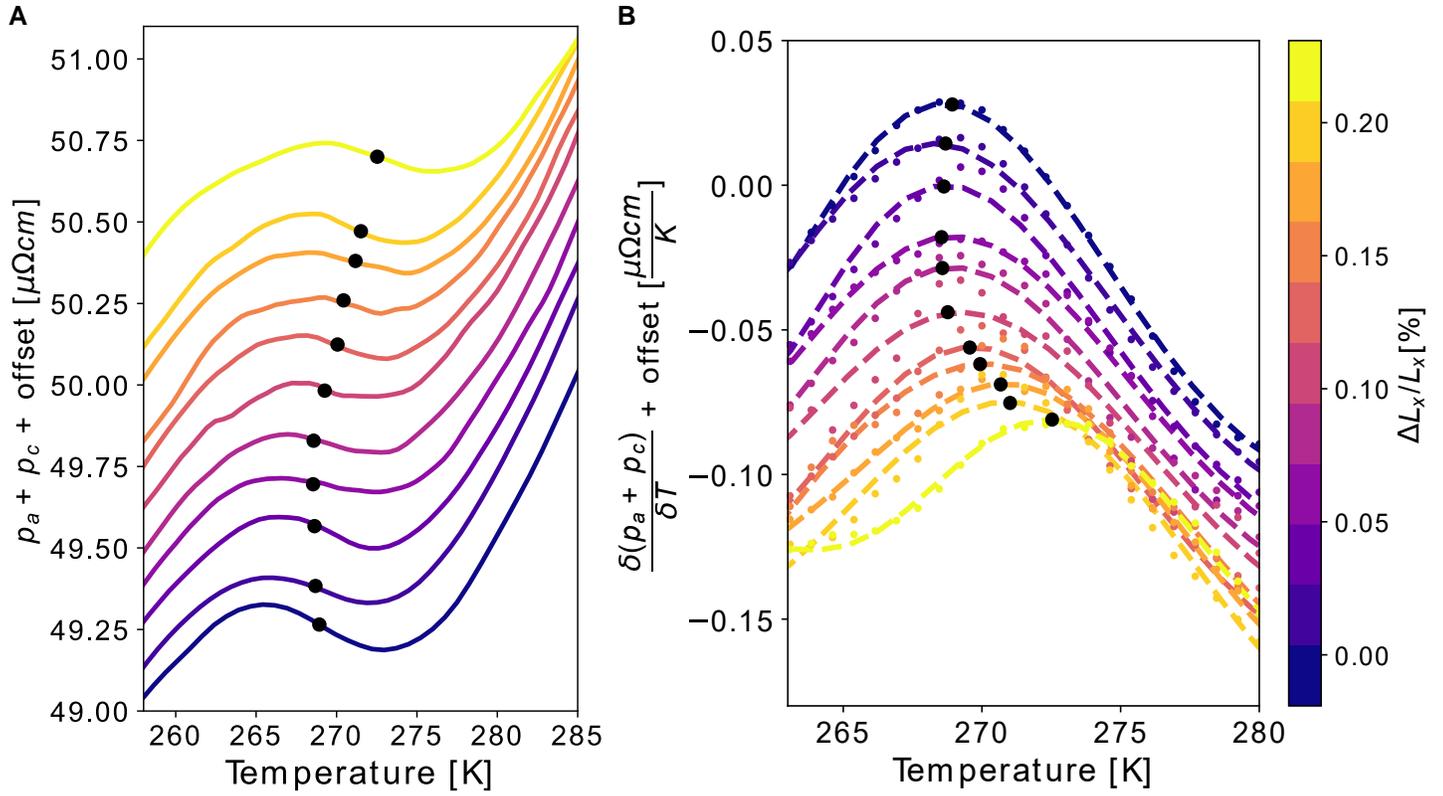

**Fig. S4.**
**Determining $T_{CDW}$:** a) $\rho_a + \rho_c$ the longitudinal resistivity plotted as a function of temperature for various offset strains indicated by the color bar. b) Temperature derivative of the longitudinal resistivity used to determine $T_{CDW}$. Black dots on both plots indicate the value of $T_{CDW}$ for that strain dataset. Identified as the peak in the first derivative, $T_{CDW}$ corresponds to the point at which the resistivity increases the fastest with cooling rather than the higher temperature point at which the resistivity stops decreasing with cooling.

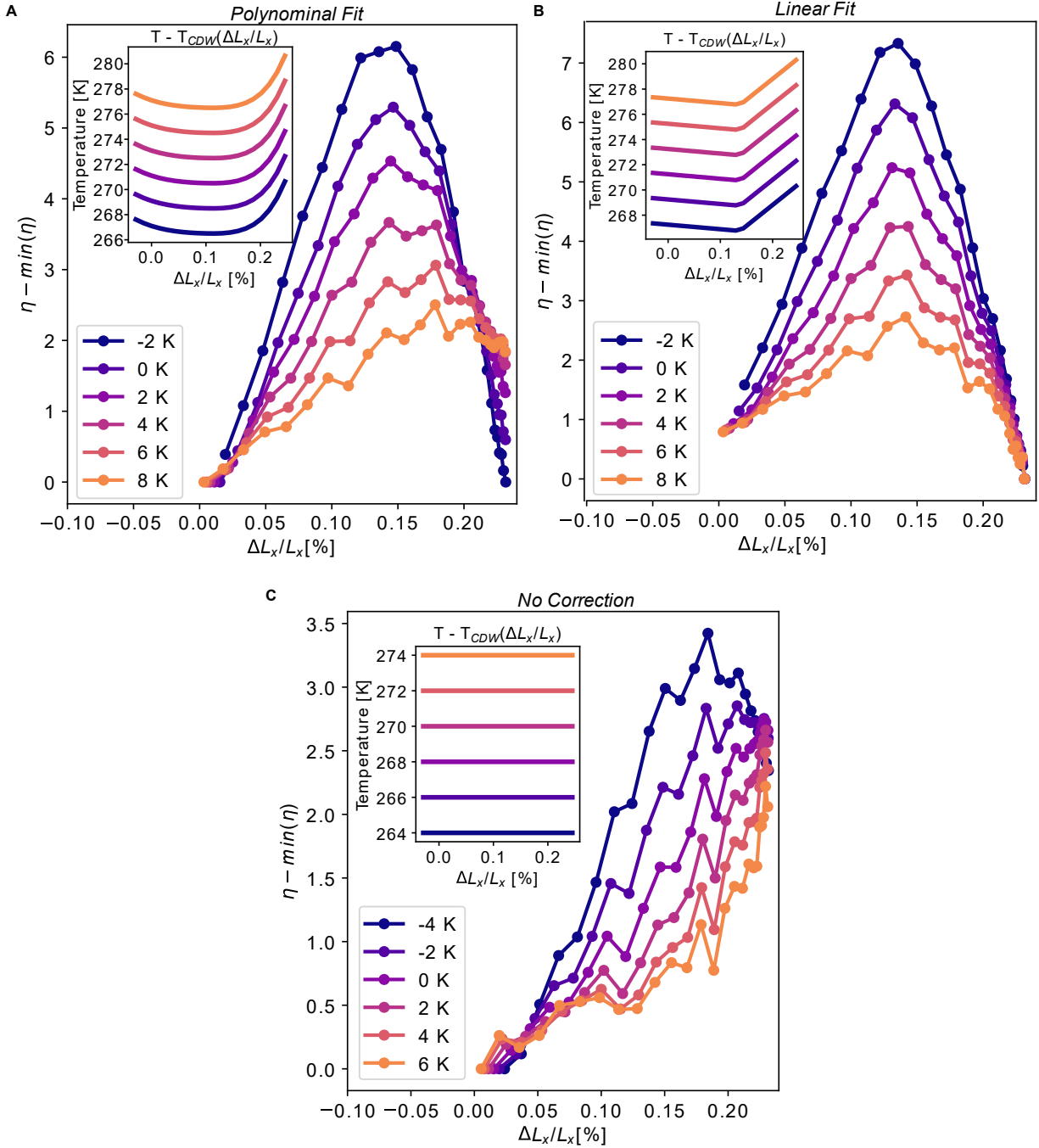

**Fig. S5.**
**Peak in η robust to changes in definition of $T_{CDW}$:** Measured values of $\eta$ at fixed values of relative temperature for different definitions of relative temperature as defined in the inset of each panel. Each data set is subtracted by its minimum value to facilitate comparison between temperatures.